\documentclass[11pt]{article}
\usepackage{moriond,epsfig}
\usepackage{graphicx,graphics}

\begin{document}
\vspace*{4cm}
\title{RECENT RESULTS FROM THE PIERRE AUGER OBSERVATORY}

\author{ A.ZECH for the PIERRE AUGER COLLABORATION }

\address{LPNHE, Universit\'{e} de Paris 6 et 7, 4 place Jussieu, 75005 Paris, aszech@lpnhe.in2p3.fr}

\maketitle
\abstracts{
The goal of the Pierre Auger Observatory is to determine the still unknown nature and origin 
of ultra-high energy cosmic rays. The study of these elusive particles probes astrophysical 
sites of particle acceleration as well as fundamental interactions at energies unattainable in 
accelerator facilities.
Auger combines two observational techniques, a large surface array and air fluorescence detectors, 
to observe the extended air showers generated in the atmosphere by cosmic rays. This hybrid 
observation yields an excellent resolution and allows for important cross-checks.
The Auger South site, located in Mendoza (Argentina), is now nearing completion, with 60\% of its
surface array and three out of its four fluorescence detectors in operation. 
First results on the energy spectrum measurement, the search for anisotropies in arrival directions
and the upper limit on the photon fraction are discussed. 
}

\section{Open questions in ultra-high energy cosmic ray physics}

Even though important advances have been made in the field of ultra-high energy
cosmic ray (UHECR) physics in the recent years, our knowledge of 
these elusive particles, which reach energies up to around 10$^{20}$ eV, is still quite limited. 
Basic questions about their nature and sources have not yet been 
answered, and the search for appropriate acceleration mechanisms is still under way.
Given their enormous energies, the list of potential sites of acceleration is limited to
the most violent astrophysical objects, such as Active Galactic Nuclei, Gamma Ray Bursts or
pulsars. An alternative explication is given by the so-called ``Top-Down'' models, in which UHECRs are the decay products of exotic super-massive particles invoked in scenarios of topological defects or superheavy dark matter. 

Another open question comes from the observation of UHECRs with energies above the threshold of
pion-production with photons of the cosmic microwave background radiation. This so-called ``GZK'' effect
is expected to cause a strong suppression of the cosmic ray flux at energies around 10$^{19.8}$ eV. Differing results on the observation of this effect in the UHECR spectra of previous experiments still await an explanation.

The aim of the Pierre Auger Observatory is to provide answers to these questions and to finally unravel
the ``mystery'' of UHECRs, thus probing energy scales that are unaccessible in the laboratory. This paper gives a short overview of our first results on the energy spectrum measurement, anisotropy searches towards the Galactic Center and the upper limit on the photon fraction. 

\section{The Pierre Auger Observatory}

The Pierre Auger Observatory (Auger) is designed to include one site in each hemisphere to allow 
full sky coverage. The Southern site is located close to Malarg\"{u}e, in the province
of Mendoza in Argentina. Once completed, Auger South will consist of four fluorescence
detector stations and 1600 surface detectors, covering an area of about 3000 km$^2$. 
At the time of writing (April 2006), three fluorescence stations with six telescopes each 
and more than 900 surface detectors are already collecting data. Another 200 surface detectors 
have been deployed in the field.

The smallness of the
steeply falling cosmic ray flux in the ultra-high energy regime (about one particle per km$^2$
per year above 10$^{19}$ eV) precludes direct detection. One observes instead the ``extensive air showers'', cascades of secondary charged particles that are triggered by interactions of UHECRs in the atmosphere. Surface detectors (SD) collect the charged particles of the air shower as it reaches the ground, whereas fluorescence detectors (FD) observe the ultraviolet fluorescence photons emitted by excited nitrogen molecules along the path of the shower. 

Auger uses water Cherenkov detectors for the SD array. These are tanks filled with 12 tons of water each and equipped with three photomultiplier tubes (PMTs), which record the Cherenkov light from particles 
entering the tank at a sampling rate of 40 MHz. The SDs record a ``footprint'' of the air shower on the
ground. Information on timing and pulse heights is used to 
reconstruct the shower geometry and thus the arrival direction of the primary UHECR. An energy estimate can be made by comparing the lateral distribution of charged particles with simulations.

The ``Fly's Eye''-type telescopes of the FD stations consist of mirrors with diameters of 3.4 meters, which collect the 
fluorescence photons and project them onto clusters of 20$\times$22 PMTs (sampling rate
10 MHz). Each mirror covers a 30$^{\circ}$$\times$30$^{\circ}$ field of view. As the shower develops in the atmosphere, its image is recorded as a track of triggered PMTs. The geometry of the shower
is reconstructed from pulse heights and trigger times of the PMTs on the track. The UHECR energy can be estimated from the total amount of fluorescence light, since air showers deposit roughly 90\% of their
initial energy into ionization of the atmosphere. 

The combination of two different types of detectors, together with its very large effective area, 
provides Auger with a unique advantage over previous experiments. Air showers that are observed
in hybrid mode can be reconstructed using the complementary information from SDs and FDs.

\section{Measurement of the UHECR energy spectrum above 10$^{18.5}$ eV}

\begin{figure}[t]
\begin{minipage}[t]{0.40\textwidth}
\mbox{}\\
\centerline{\includegraphics[width=\textwidth]{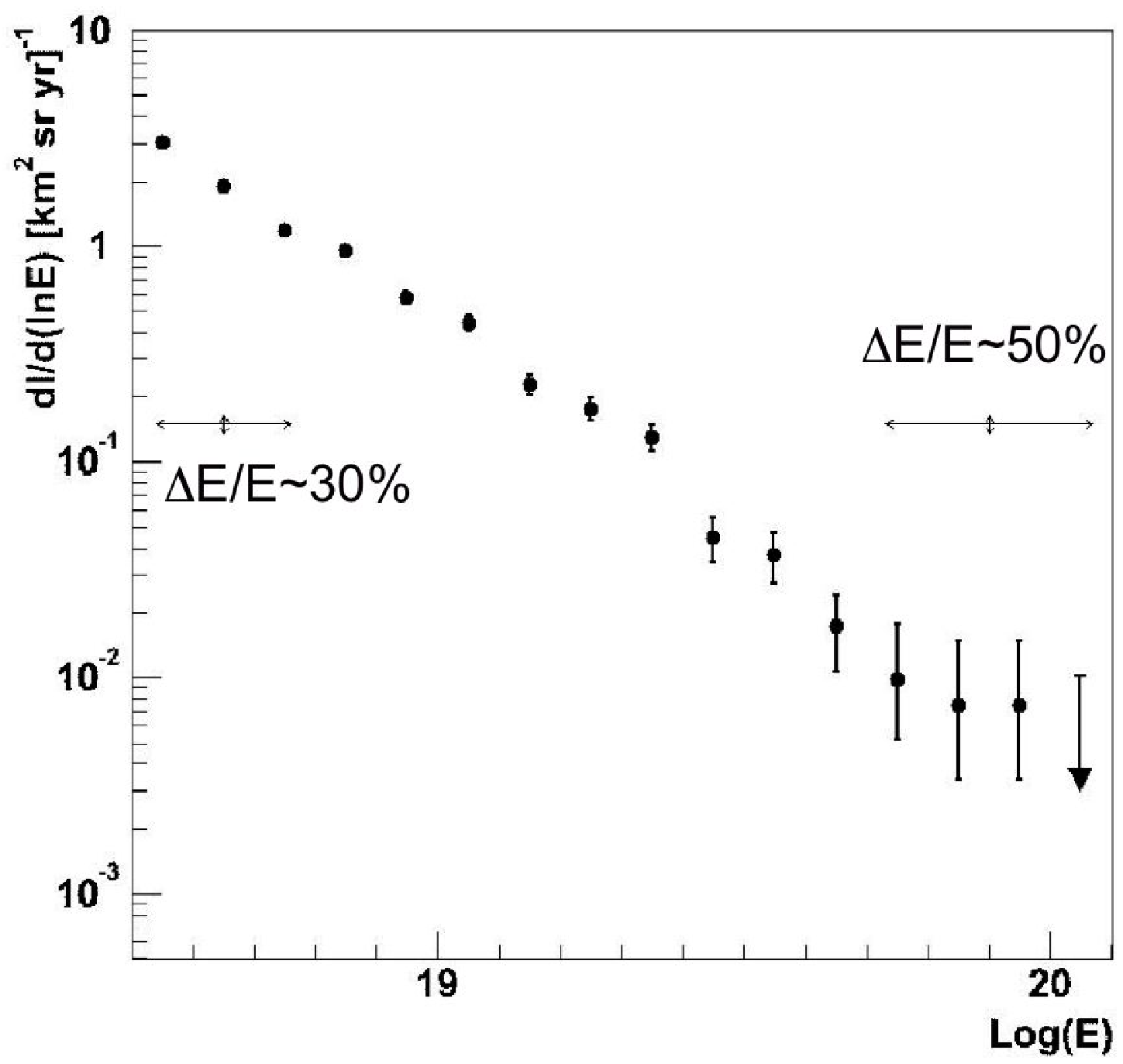}}
\caption[energy spectrum] {\raggedright Energy spectrum (flux $\times$ energy) measured by Auger. Error bars on the data points are statistical uncertainties. The two horizontal error bars indicate systematic uncertainties in the energy.}
\label{fig:augerspec}    
\end{minipage}
\hfill
\begin{minipage}[t]{0.56\textwidth}
\mbox{}\\
\centerline{\includegraphics[width=\textwidth]{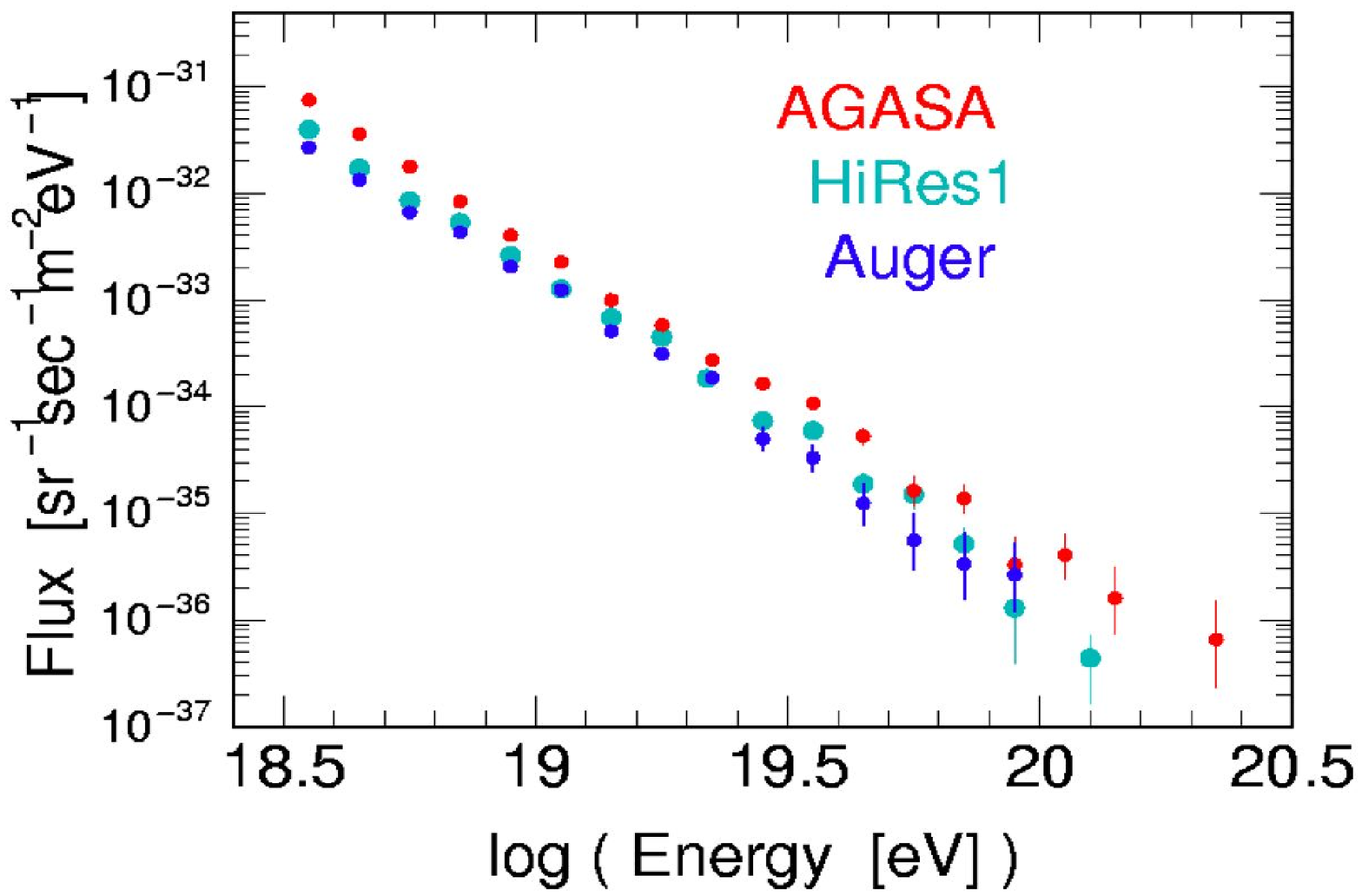}}
  \caption[energy spectra Auger, AGASA, HiRes-I]{\raggedright The energy spectrum measured by Auger (lowest points) in comparison with the AGASA (uppermost points)~\cite{agasa_spec} and HiRes-I (light blue/grey points)~\cite{hires_spec} results.}
\label{fig:augerspeccomp}
\end{minipage}
\end{figure}

The large aperture of the Auger surface array will allow for the first time an observation
of the UHECR spectrum around the ``GZK'' energy threshold with good statistics. If the
expected flux suppression is found in the spectrum, the measurement of the exact shape of 
this feature will help us to distinguish between different models of source distributions. 

Moreover, its hybrid design will allow Auger to resolve the discrepancy between previous spectrum measurements that were based on the different techniques. The surface array AGASA measured a continuation of the 
spectrum up to the highest energies~\cite{agasa_spec}, which has caused much speculation on new physics at the
UHECR energy scale. The fluorescence detectors of the HiRes experiment~\cite{hires_spec}, on the other hand,  observed a break in the spectrum at the expected energy of 10$^{19.8}$ eV.
The combination of surface and fluorescence detectors in a single experiment will
give us valuable insights in the different systematics of the two techniques and allow us to explain 
this discrepancy. Studies on this issue are currently in progress.

Our first measurement of the UHECR energy spectrum combines the advantages of the two techniques
employed in the Pierre Auger Observatory. The 100\% duty cycle of the SD provides us with sufficient statistics, even though
the analyzed data (1 January 2004 through 5 June 2005) correspond to less than one third of the amount we anticipate for one year of data collecting, once the Auger South array will be completed. At the same time, we exploit the fluorescence information from a subset of showers observed in hybrid mode to determine the absolute energy scale. The energy estimate uses the signal size at a radius of 1000 m from the shower core ( ``S(1000)'' ), which is determined from a fit to the lateral distribution
of signal sizes from all the tanks triggered by an air shower. The ``Constant Intensity Cut'' method~\cite{auger_spec} is used to re-scale values from different shower inclinations. S(1000) is almost linearly proportional to the energy of the primary particle. We extract the conversion factor that relates S(1000) to the energy from the hybrid events by use of the very good energy reconstruction based on the FD information. This reduces significantly our 
dependence on air shower models and on assumptions of the UHECR composition, compared to
previous surface array experiments.

Figure~\ref{fig:augerspec} shows the energy spectrum (differential flux $\times$ energy). 
The two horizontal error bars indicate the systematic
uncertainty in the energy determination. The increase of this uncertainty with higher energies comes from the S(1000)-to-energy conversion factor, which still suffers from the limited statistics in the hybrid data set and will improve quickly as more data are being collected. The two small
vertical error bars indicate the systematic uncertainty in the estimate of the exposure of the SD array.  Since the SDs are 100\% efficient at energies above 10$^{18.5}$ eV, the exposure of the array depends only on its geometry and on the live-time of the detectors. The cumulative exposure of the data included in the spectrum adds up to 1750 km$^2$ sr yr. This is already larger than the total exposure of the AGASA experiment.

A comparison with the results from AGASA and HiRes-I is shown in Figure~\ref{fig:augerspeccomp}. Our data points are about 10\% below the HiRes points. No 
``super-GZK'' events have been found so far. However, at this early stage in our analysis
we cannot draw any conclusions about the existence of the GZK feature. The uncertainties in our
spectrum are still too large and systematic effects between the analyses based on FD and SD
data need to be studied in detail first.

\section{Anisotropy searches}

We have analyzed SD data from the same period as described above to search for anisotropies in cosmic ray
arrival directions near the Galactic Center, Galactic Plane and Supergalactic plane. Two previous experiments, AGASA~\cite{agasa_gc} and SUGAR~\cite{sugar_gc}, claim significant excesses in the cosmic ray flux near to the
Galactic Center at energies of 1 - 2.5 EeV and 0.8 - 3.2 EeV, respectively. 
The detection of a TeV $\gamma$ ray source in the Galactic Center and of diffuse $\gamma$ ray emission along the Galactic Plane by the HESS collaboration~\cite{hess} has provided additional indirect evidence for the acceleration of cosmic rays at these locations, albeit at much lower energies.

The statistics of our first data set in the Galactic Center region are already larger than that of any previous experiment. For the SD data, the angular resolution is better than 2.2$^{\circ}$. We have
searched for excesses at the angular scale of our resolution and at the scales and energies used by the previous experiments.

\begin{figure}[t]
\begin{center}
\includegraphics[height=25pc]{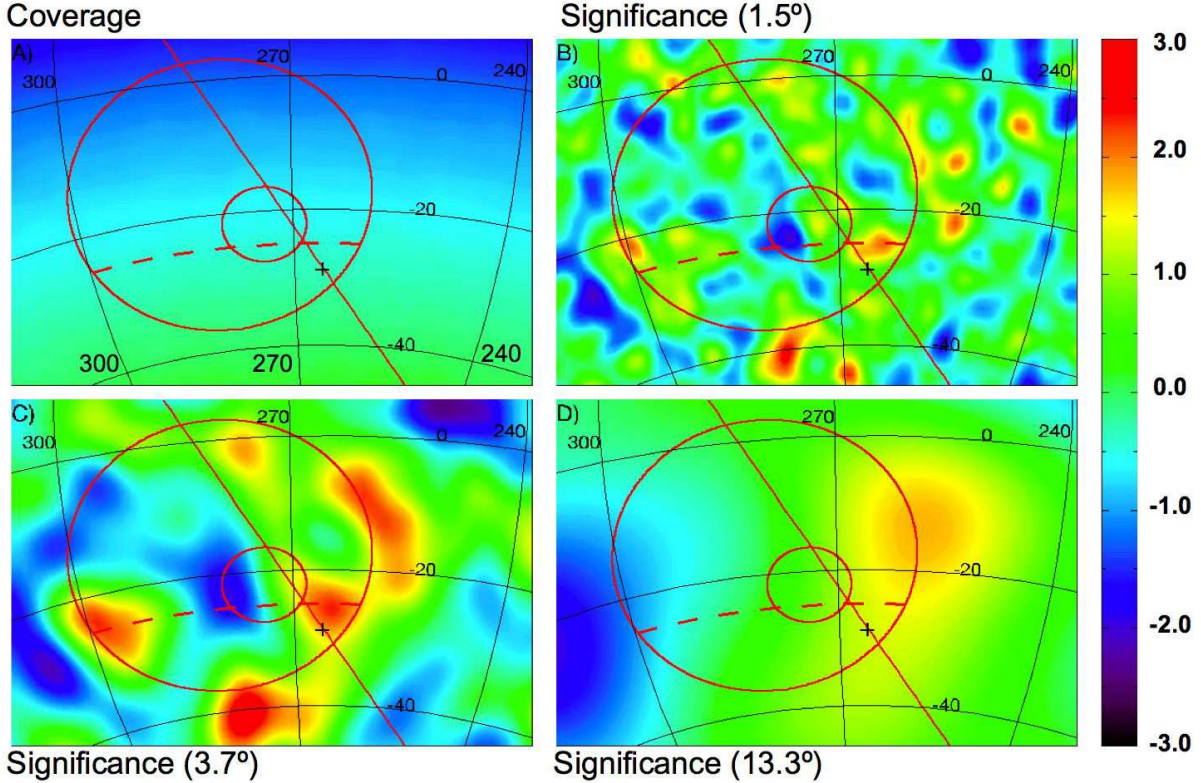}
\end{center}
\caption{Results from anisotropy searches in the direction of the Galactic Center (cross).
See text for explanation.}
\label{fig:galcenter}
\vspace{- 0.1 cm}
\end{figure}

Figure~\ref{fig:galcenter} combines the results of this search. Panel A shows the Auger coverage map in the vicinity of the Galactic Center (cross). The excess regions observed by AGASA (large circle) and SUGAR (small circle) are indicated, as well as the Galactic Plane (line) and the lower limit of the AGASA
field of view (dashed line). The coverage map contains the expected distribution of event
densities for purely isotropic sources. It has been constructed using the ``shuffling'' technique~\cite{auger_gc}$^,$~\cite{auger_map}. The distribution of observed events is compared to this map in order to determine a significance map. 

Panel B is the significance map at the scale of the Auger angular resolution and in the SUGAR energy range. Different colors (or shades of grey) indicate the significance of excesses in standard deviations. Panels C and D show the significances at the SUGAR and AGASA angular scales, respectively. Energy ranges have 
been matched to the intervals quoted in the two experiments. In separate studies, the energy intervals
have also been adjusted to take possible differences in the calibration into account.

No significant excesses have been found in the Auger data. With our data set, which contains roughly three times the statistics
of AGASA and more than ten times the statistics of SUGAR, we would expect a significance of 7.5$\sigma$ and 10.5$\sigma$, respectively, given the excesses seen by those experiments. However, the fluctuations we see are consistent with an isotropic flux. 

This result has also been confirmed with hybrid data, which have a much better angular resolution
of 0.6$^{\circ}$. Similar searches along the Galactic Plane and Supergalactic Plane 
yield results consistent with isotropy. A separate blind search and a prescription search did not find
any significant excesses either~\cite{auger_anis}.

\section{Upper limit on the photon fraction above 10$^{19}$ eV}

With the hybrid data collected by Auger South from January 2004 to April 2005, we have
determined an upper limit for the photon fraction at primary energies above 10$^{19}$ eV. 
The importance of the photon fraction lies in the fact that ``Top-Down'' models predict a 
considerable proportion of photons among the generated particles. The measured photon
flux is thus a valuable indicator of these non-acceleration models. 

Our method to distinguish between hadrons and photons in the hybrid data set exploits the
information on the longitudinal profile of the air shower that is provided by the FD observation.
The atmospheric depth at the shower maximum is known as X$_{max}$ and is commonly used as a discriminant observable
for the cosmic ray composition. This exploits the fact that lighter nuclei penetrate on the average more deeply into the atmosphere.
At ultra-high energies, showers initiated by photons develop significantly deeper in the atmosphere than hadronic showers, yielding X$_{max}$ values well above those of proton showers.

\begin{figure}[t]
\begin{minipage}[t]{0.48\textwidth}
\mbox{}\\
\centerline{\includegraphics[width=\textwidth]{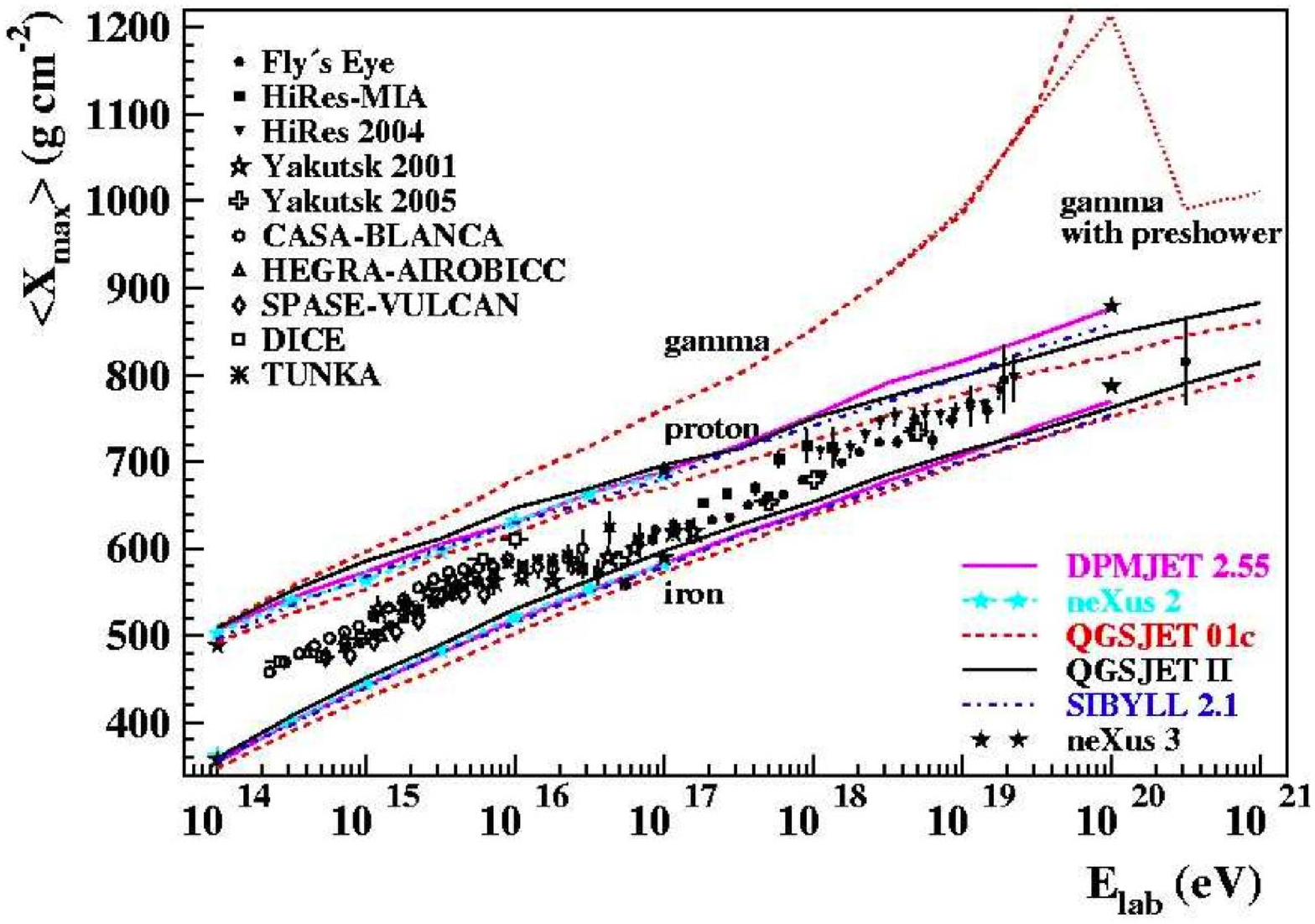}}
 \caption[Xmax vs. energy] {\raggedright Mean X$_{max}$ versus energy from model predictions (lines) and previous experiments.}
\label{fig:xmaxcomp}    
\end{minipage}
\hfill
\begin{minipage}[t]{0.48\textwidth}
\vspace{0.9cm}
\mbox{}\\
\centerline{\includegraphics[width=\textwidth]{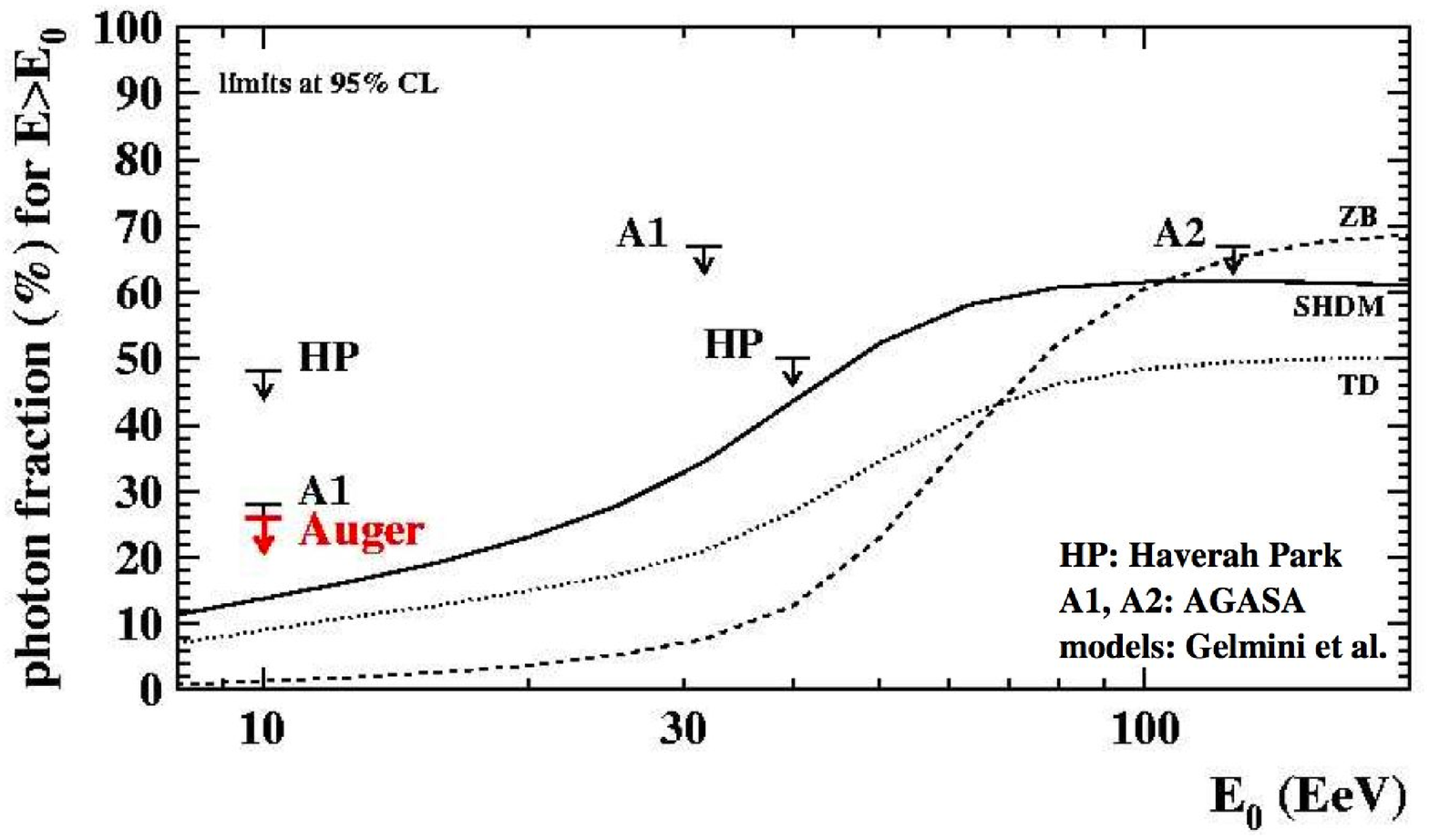}}
  \caption[photon fraction upper limits]{\raggedright Upper limits on the photon fraction from AGASA, Haverah Park and Auger. Predictions for different ``Top-Down'' models (Z Bursts, Super Heavy Dark Matter, Topological Defects) are included. }
\label{fig:upperlimit}
\end{minipage}
\end{figure}

Figure~\ref{fig:xmaxcomp} shows model predictions for proton, iron and photon cosmic rays in comparison with measurements by previous experiments. As can be seen, the separation in the predictions for hadrons and photons at 10$^{19}$ eV is larger than 200 g/cm$^2$, whereas a very conservative estimate of the systematic uncertainty in our X$_{max}$ determination yields $\simeq$ 40 g/cm$^2$.

In our set of hybrid events, no candidate for a primary photon was found~\cite{auger_photon}. By comparing the 
observed X$_{max}$ of each event to predictions from photon shower simulations, we have derived
an upper limit on the photon fraction from our data. At a confidence level of 95 \%, we find an upper 
limit of 26\% above 10$^{19}$ eV. This result is shown in Figure~\ref{fig:upperlimit} together with 
upper limits from the AGASA~\cite{agasa_photons} and Haverah Park~\cite{hp_photons} ground 
arrays. Our result confirms and improves the existing limits above 10$^{19}$ eV. The figure includes
also predictions for the photon fraction for three ``Top-Down'' models. With the limited statistics of the analyzed hybrid data, we are not yet able to put constraints on any of the models or to obtain limits at higher energies. However, our rapidly increasing statistics will allow us to soon improve this result significantly.

In a separate analysis based on SD data and using the ``rise time'' and the curvature of the shower
front as discriminant observables~\cite{auger_photon}, we will be sensitive to a 
significantly lower fraction of gamma rays, thus examining the question of 
a photon origin much more rigorously.

\section{Conclusions}

The first results from the Pierre Auger Observatory demonstrate the power of its large aperture 
and hybrid design. Even though stable operations began only in January 2004 and the analyzed
data period is marked by an increasing aperture, the collected data statistics yield already important
new insights.

The energy spectrum above 10$^{18.5}$ eV has been determined with SD data, with a calibration using the fluorescence technique. Our spectrum has a normalization closer to the HiRes measurement, but it is too early to draw conclusions concerning the GZK feature. 

Our SD data set is not consistent with the anisotropies towards the Galactic Center observed by AGASA and SUGAR, although the collected statistics are already several factors larger than in those experiments. 
No excesses of events from the Galactic Plane or Supergalactic Plane could be observed. In the energy ranges around 10$^{18}$ eV,  the cosmic ray flux has been found to be isotropic.

With our hybrid data set, we could put an upper limit of 26\% on the photon fraction above 10$^{19}$ eV at a 95\% confidence level. This confirms and improves previous results by AGASA and Haverah Park. 

The last FD station of Auger South is currently under construction and less than one third of the surface detectors remain to be deployed. While the Southern site is nearing its completion, planning for the Northern Site in Colorado is in progress. 

\section*{Acknowledgments}
I wish to acknowledge the Marie-Curie fellowship program of the European Union
for financial support of my research work within the Pierre Auger Collaboration.

\end{document}